\documentclass[doublecol]{epl2} 
\usepackage{amsmath,amssymb,amsfonts}
\usepackage{graphicx}
\graphicspath{{./}{../submission/}}
\newcommand{\E}{\mathbb E}
\title{Scaling of Information in Turbulence}
\author{Carlos Granero-Belinch{\'o}n, Stephane G. Roux, Nicolas B. Garnier}
\institute{  
Univ Lyon, Ens de Lyon, Univ Claude Bernard, CNRS UMR 5672, Laboratoire de Physique, F-69342 Lyon, France
}


\abstract{
We propose  a new perspective on Turbulence using Information Theory. We compute the entropy rate of a turbulent velocity signal and we particularly focus on its dependence on the scale. 
We first report how the entropy rate is able to describe the distribution of information amongst scales, and how one can use it to isolate the injection, inertial and dissipative ranges, in perfect agreement with 
the Batchelor model and with a fractional Brownian motion (fBM) model. We provide analytical derivations of the entropy rate scalings in these two models.
In a second stage, we design a conditioning procedure in order to finely probe the asymmetries in the statistics that are responsible for the energy cascade. 
Our approach is very generic and can be applied to any multiscale complex system.
}

\begin{document}

\maketitle

\section{Introduction}

\subsection{Turbulence}

Despite many progresses in understanding Turbulence during the last century, a lot of questions remain unanswered.
In 1921, Richardson depicted Turbulence as a hierarchy of whirls of different sizes, with an energy cascade from larger eddies down to smaller ones~\cite{Richardson1921}.

This description inspired Kolmogorov theory of fully developed turbulence~\cite{Kolmogorov1941a,Kolmogorov1941b, Kolmogorov1941c}. The experimental observation of Intermittency~(see, {\em e.g.}, \cite{F.Anselmet1984}) 
led to corrections to the K41 theory \cite{Kolmogorov1962} and to a description of the multifractal nature of Turbulence signals~\cite{Frisch1985, Frisch1995}.

We propose in this article an Information Theory perspective on Turbulence: we first describe the distribution of information (in the Shannon's sense) over scales in a turbulent velocity signal and then show and 
explain its close connexion with the energy scaling. In order to explore the existence of an energy cascade in Turbulence via Information Theory, we define a conditioning procedure of the signal with respect to 
the sign of its increments, and show that this allows for a very fine exploration of the asymmetry of the distribution and hence the inference of the existence of a finite skewness. The methods presented here are 
of general interest in the study of complex dynamical signals. 

\subsection{Information Theory}
Since Shannon's pioneering work~\cite{Shannon1948}, Information Theory (IT) has been developed to analyse the complexity or disorganization of signals. Although strongly connected with correlations, entropy and 
other quantities subsequently defined in IT do not rely on two points correlations only, but on joint probabilities.
These quantities are then in theory able to handle not only higher order moments of any Probability Density Function (PDF), but also all possible non-linear relationships between signals. In the present article, we mainly deal with entropy rate, and mutual informations, although many other more elaborated extensions can 
be considered.

\subsection{Previous studies and difficulties}
Up to now, very few research has been devoted to applying IT in the analysis of Turbulent signals~\cite{Ikeda1989,R.T.Cerbus,R.T.Cerbus2013,MassimoMaterassi2014}. 
This has probably two main origins. 
Firstly, all quantities in IT usually require a large number of samples or data points to behave properly, with small enough bias and variance. 
Secondly, correlations ---~within a single signal or between signals~--- impact the estimation of the probability density functions.
Usually, one is not interested in these effects of correlations and uses the Theiler prescription~\cite{Theiler1986} to discard them.

In this paper, we analyse a Turbulent velocity signal using Information Theory. Our "key" tool is the entropy rate, and we focus on its dependence on the scale. Doing so, we do not care about the Theiler correction; 
on the contrary, we want to explore all domains where correlations are important and which contain useful information. 
In a first stage, we report how the entropy rate is able to describe the distribution of information amongst scales. 
In a second stage, we design a conditioning procedure to extract asymmetries in the statistics that are responsible for an energy cascade.

\section{Information measurements}

\subsection{Definitions}

Shannon entropy of a signal $X$ that takes its values in the vector space $S$ is a functional of the PDF $p(x)$ of the signal $X$ on $S$~\cite{Shannon1948}:

\begin{equation*}
H(X)=- \int_{S} p(x)\log p(x) {\rm d}x
\end{equation*}

We focus here on real, unidimensional signals, so $S$ is the set of real numbers. One important property of entropy is that it does not depend on the first moment of the PDF. Therefore, in the remainder of this 
article, we are working with centered signals.

In order to study the dynamics of the signal, we use a delay-embedding procedure \cite{Takens1981} to construct the $m$ dimensional signal $X^{(m,\tau)}_{l}$ from the signal $X$ by defining for each time $t=t_0+l$ 
the vector:
\begin{equation*}
\textbf{x}_t^{(m)}=(x(t),x(t-\tau),x(t-2\tau),..., x(t-(m-1)\tau))
\end{equation*}
\noindent where $m$ is the embedding dimension and $\tau$ is a time delay that we adjust. Here and in the following we omit the index $l$ if $l=0$, and $(m,\tau)$ if $m=1$. 

\medskip

The entropy rate measures the additional information that is brought by the $(m+1)$-th dimension when the $(m)$ previous ones are known:
\begin{align}
h_\tau^{(m)}(X) &\equiv  H(X_{\tau} | X^{(m,\tau)}) \nonumber \\
&=  H(X^{(m+1,\tau)}) - H(X^{(m,\tau)}) \label{eq:def:h:diff} \\
&=  H(X) - I(X_{\tau}, X^{(m,\tau)}) \label{eq:def:h:MI}
\end{align}

The last equation involves the mutual information (MI)~\cite{Shannon1949} which is defined between two embedded signals $X^{(m)}$ and $Y^{(p)}$ as

\begin{align*}
I(X^{(m,\tau)},Y^{(p,\tau)}) & = \\
H(X^{(m,\tau)}) &+ H(Y^{(p,\tau)}) - H(X^{(m,\tau)},Y^{(p,\tau)}).
 \label{eq:def:MI}
\end{align*}

In eq.(\ref{eq:def:h:MI}), the MI is computed between two specific signals $X_\tau=X ^{(1,\tau)}_{\tau}$ and $X^{(m,\tau)}$ built from the signal $X$ such that the concatenation of the two signals is nothing but the $(m+1)$-embedded signal $X$:  
$(X^{(1,\tau)}_{\tau}, X^{(m,\tau)})=X_{\tau}^{(m+1,\tau)}$.

The entropy rate measures the "new" information in $x(t+\tau)$ that is not in $x_t^{(m)}$. In this perspective, an increase of $m$-points correlations, so an increase of 
the information contained in one point about its neighbors,
can be seen as an increase of organization and therefore a decrease of the entropy rate.
From eq.(\ref{eq:def:h:diff}), we see that the entropy rate can be negative if the information contained in the embedded signal $X^{(m,\tau)}$ is larger than the information contained in the embedded signal $X^{(m+1,\tau)}$; this situation is expected when correlations are very strong. 
From eq.(\ref{eq:def:h:MI}) and noting that the Mutual Information is always positive, the sign of the entropy rate depends on the magnitude of the entropy $H(X^{(m,\tau)})$.
If the signal $X$ has a continuous support, the entropy ---~which depends on the standard deviation of the signal~---
can be negative.

The entropy rate $h_\tau^{(m)}$ depends on the time scale $\tau$ used in the embedding process. Contrary to most of the literature on Information Theory, we are interested here in the dependence of the entropy 
rate in $\tau$, and we are varying $\tau$ from its smallest value ---~the sampling period $dt$~--- up to some large time scale. This is motivated by the connexion that we discuss below between the embedding process, 
and the "traditional" definition of increments, both using the time scale $\tau$.

\medskip

The entropy rate $h_\tau^{(m)}$ is related not only to 2-point correlation functions, but also to higher order statistics, in particular if $m>2$.
We report below how to estimate the entropy rate for an experimental turbulent signal, and describe its dependance on the time scale.

\subsection{Estimating the entropy rate}

A natural estimation of the entropy rate is obtained by computing the entropy of two successively embedded version of the signal $X$, for embedding dimension $m$ and $m+1$, and 
then substracting them according to eq.(\ref{eq:def:h:diff}). This has unfortunately several drawbacks. When the embedding dimension $m$ is increased, the bias of the measure increases as well; this is known as the 
curse of dimensionality. One has therefore to make a trade-off between on one hand a larger value of $m$ for a theoretically better estimation of the entropy rate, and on the other hand a smaller value of $m$ to avoid bias. 
Moreover ---~and whatever the choice of $m$ is~--- the expected bias for $H(X^{(m,\tau)})$ and $H(X^{(m+1,\tau)})$ is {\em a priori} different and results in a larger bias for the entropy rate than for any of the two 
entropies considered separately.

Fortunately, one can exploit the expression (\ref{eq:def:h:MI}) of $h^{(m)}_{\tau}$ based on Mutual Information, for which Kraskov et al. \cite{Grassberger2004} provided a $k$-nearest neighbors estimator which has several good properties. 
Amongst them is not only a small Mean squared error even for moderate signal sizes $N$, but also some build-in cancellation of the bias difference from the two arguments $X^{(m,\tau)}$ and $Y^{(p,\tau)}$. It then suffices to subtract the entropy 
$H$ for $m=1$, which is relatively easy to estimate. Several algorithms exist, and we use the $k$-nearest neighbors estimate from Kozachenko and Leonenko~\cite{L.Kozachenko1987}.

\section{Turbulence signal analysis}

The data we are analysing is a temporal measurement (sampled at frequency $f_s$=25 kHz) of velocity in a grid turbulence experiment in the ONERA wind tunnel in Modane~\cite{Castaing1998}. 
The Taylor-scale based Reynolds number is about $R{\rm e}=2700$ and the turbulence rate is about 8\%; the Kolmogorov $k^{-5/3}$ law for the energy spectrum holds on an inertial range of approximately three time decades (Figure~\ref{fig:results:h}(a)). 
In the remainder of this article, we implicitly use the Taylor hypothesis~\cite{Frisch1995} and consider our 
temporal data as representing spatial fluctuations of the longitudinal velocity. The integral scales, inertial range and dissipative range that we describe below in terms of time scales or ranges are to be 
understood as corresponding to spatial time scales or ranges. The proportionality factor is the mean velocity $\langle v \rangle$ of the signal ($\langle v \rangle=$ 20.5 m/s here).

The probability density function of the data is almost Gaussian although there is some visible asymmetry (the skewness is about $0.175\pm0.001$).

\subsection{Entropy rate} 

\begin{figure}
\includegraphics[width=1\linewidth]{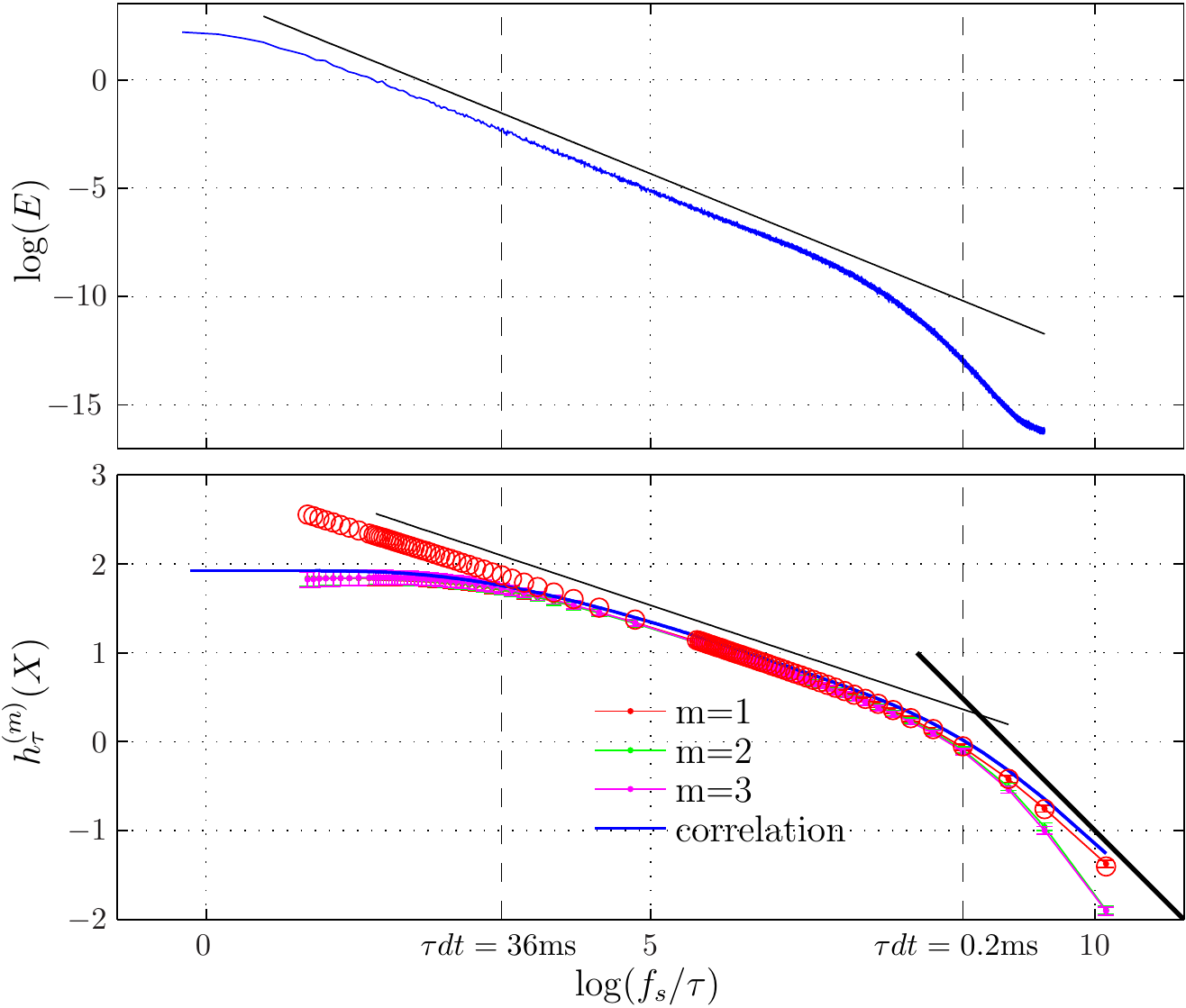}
\caption{Top: Power Spectral Density of the experimental signal. The straight line corresponds to the K41 scaling. 
Bottom: entropy rate $h^{(m)}_\tau$, as a function of the scale ($f=f_s/\tau$) for different embedding dimensions $m\in\{1,2,3\}$. The blue curve corresponds to eq.(\ref{eq:h:corr}) using the autocorrelation function.
The thin straight line is a line of slope -1/3, the thick straight line has a slope -1.}
\label{fig:results:h}
\end{figure}

In the following, the time-lag is written $\tau dt$ where $\tau$ is a non-dimensional integer, and $dt=\frac{1}{f_s}$. 
Thanks to the Taylor hypothesis $\tau dx$ can be interpreted as a spatial scale, with $dx=\langle v \rangle dt$.
We estimate the entropy rate  $h_\tau^{(m)}(X)$ using eq.(\ref{eq:def:h:MI}) with vectors $X_{\tau}$ and $X^{(m,\tau)}$ of sizes $(1\times N$) and $(m\times N)$ where $N=2^{17}$ is kept constant for all $\tau$.
The entropy rate is computed, and then averaged, over $195$ independent samples. 
Our results are presented in Fig.\ref{fig:results:h} as a function of $\log(f_s/\tau)$.
Three regions are observed: 
above $36$ms are the integral scales, below $\sim 0.18$ms is the dissipative range, and in-between them is the inertial range. 
At larger (integral) scales, the auto-correlation function vanishes and the entropy rate is equal to the entropy $H$ of the signal $(m=1)$ which depends only on one-point statistics. We can interpret this first 
result as indicating that the integral scales, where energy is injected and Turbulence is generated, are the most disorganized. In the inertial range, the entropy rate decreases almost linearly with a slope close 
to -1/3, represented by a straight line in Fig.~\ref{fig:results:h}. As the time ---~or space~--- scale is decreased, the velocity field appears more and more organized in the sense that the amount of "new" information brought 
to one point by another point at a distance $\tau$ decreases with $\tau$. In the dissipative range, the entropy rate decreases faster and faster, as the dissipation become stronger and stronger. We interpret this as 
a consequence of the velocity field being more and more regular, as the scale decreases.

The entropy rate enlightens differently than the Power Spectrum the separation between the different domains, as can be seen in Fig.~\ref{fig:results:h}. The integral and Kolmogorov scales shown in this figure have been obtained 
using the Batchelor model for fully developed turbulence, as detailed below.

\subsection{Robustness} We checked that our estimation of the entropy rate does not depend on the number of 
neighbors used in the k-nn search algorithm ($k=5$ and $10$) nor on the  sample size $N$ ($N=2^{16}, 2^{17}$ and $2^{19}$). We measured that for a fixed set $(k,N)$ the standard deviation of the entropy rate is 
much smaller than the standard deviations of $H(X)$ and $I(X_{\tau}, X^{(m,\tau)})$ considered separately.  
The standard deviation of the estimation is of the order of $0.03$ for small $\tau$ and increase to $0.08$ for large $\tau$.

\subsection{Effect of the embedding}
We varied the embedding dimension $m$ and observed a dependance of the entropy rate in the dissipative range only, for scales $\tau dt$ smaller than $0.18$ms. Going deeper and deeper in this range, the signal is more 
and more continuous so the knowledge of an increasing number $m$ of points in the past (separated by the scale $\tau$) decreases significantly the "new" information brought by a $(m+1)$th point in the 
future~(Figure~\ref{fig:results:h}). But for $m>2$, there is no measurable evolution anymore, even in the dissipative range.
 
\subsection{Entropy rate and autocorrelation} 
If the statistics of the signal are Gaussian, the entropy rate can be expressed as
\begin{equation}
h_\tau^{(1)}(X)= H(X) + \frac{1}{2}\log(1-c(\tau)^{2})
\label{eq:h:corr}
\end{equation}
where $c(\tau)$ stands for the normalized  autocorrelation function ($c(\tau=0)=1$) and $H(X)=\frac{1}{2}\log(2\pi e \sigma^2)$ with $\sigma$ denoting the standard deviation of $X$, is independent of $\tau$. 
Eq.(\ref{eq:h:corr}) takes into account the two-point correlations only but it gives surprisingly good results when applied to Turbulence data. 
Note that from formula (\ref{eq:h:corr}) it is obvious that strong correlations can lead to a negative entropy rate.
The entropy rate obtained using the auto-correlation function $c(\tau)$ and eq.(\ref{eq:h:corr}) is presented as a blue line in Fig.\ref{fig:results:h}(b). It is almost identical to the estimate for $m=1$. 
This result is not surprising because the Turbulence signal has almost Gaussian statistics, which implies that higher order correlations are small and difficult to observe. In this case, eq.(\ref{eq:h:corr}) 
---~relating the entropy rate and the auto-correlation function $c(\tau)^{2}$~--- holds, and the signal is described equivalently whether using the auto-correlation function, the second order structure function, 
or the entropy rate with $m=1$ which depends explicitly on two-point statistics. For $m>1$, the entropy rate depends {\em a priori} on the $(m+1)$-point correlation structure, but these higher order correlations 
are barely observable in the Turbulence signal, which is confirmed by our analysis in the integral and inertial ranges. Only in the dissipative range a difference with the autocorrelation estimate appears.
This behavior at the smallest scales may not be due to Turbulence itself; rather it may originate from the experimental processing of the signal.

We can use eq.(\ref{eq:h:corr}) to explore the inertial range and its boundaries by using the Batchelor model for fully developed turbulence~\cite{Batchelor1951} which provides a model of $S_2$, the second-order structure function:
\begin{equation}
S_2(\tau)=\frac{(\tau dx/L)^{2/3}}{\left( 1+\left( \eta_k/\tau dx \right)^{2}\right)^{2/3}}\,.
\label{eq:Batchelor}
\end{equation}
$L$ is the integral scale and $\eta_{k}$ the dissipation scale, and $dx=\langle v \rangle dt$ from the Taylor hypothesis. 
Eq.(\ref{eq:Batchelor}) imposes on the second-order structure function to have a slope $2/3$ in the inertial range and $2$ in the dissipative range. Noting that $S_2(\tau)=1-c(\tau)$ and using eq.(\ref{eq:h:corr}), one derives for the entropy rate $h^{(1)}(\tau)$ two linear behaviors in $\log(1/\tau)$: one with a slope $-1/3$ in the inertial region and another one with a slope of $-1$ in the dissipative range, both in perfect agreement with our measurements.
Fitting $S_2$, the structure function, or fitting the entropy rate give the same estimates of the scales $L$ and $\eta_{k}$; these are represented as vertical dashed lines in Fig.~\ref{fig:results:h}.

\subsection{Fractional Brownian Motion}

Fractional Brownian motion (hereafter fBm) is a continuous-time random process proposed by Mandelbrot and  Van Ness~\cite{M68} in 1968, which quickly became a major tool in various fields where concepts of self-similarity and  
long-range dependence are relevant. In order to reproduce the K41 scaling of turbulence, we consider here a fBm $B$ with an Hurst exponent ${\cal H}=1/3$~\cite{M68,Kolmogorov1941a}. Although this signal is non stationary, its increments are Gaussian and stationary; moreover, it has the same correlation structures as a turbulent velocity signal in the inertial range: its Power Spectral Density has a power law with exponent $-5/3$. Using $dt=1$ for this synthetic signal, the non-stationary covariance structure $\E \{B(t)B(t+\tau) \} =  \sigma^2_0 c(t,\tau)$ is given by
\begin{align*}
\sigma^2_0 c(t,\tau) =
\frac{\sigma^2_0}{2} \left[t^{2H}+(t+\tau)^{2H}-|t-(t+\tau)|^{2H}\right] 
\end{align*}
The pre-factor $\sigma_0$ is a normalization constant. 
The fBm having Gaussian statistics, we can write its entropy as
\begin{equation}
H_{\rm FBM}^{(m)}\equiv H(B^{(m,\tau)})=\frac{1}{2} \log((2 \pi e)^m |\Sigma|)
\end{equation}
where $\Sigma$ is the $m\times m$ covariance matrix with coefficients $\Sigma_{i,j}=c(t_i,t_j)$.
$\sigma^2_0$ is a scaling factor independent of time.
The fBm is non stationary, but if we consider a signal of finite temporal extension $T$, we obtain analytically  for $m=1$ :
\begin{align}
h_\tau^{(1)}(B) &\approx {\cal H}\log\left(\tau \right) + \frac{1}{2}\log\left(2\pi e \sigma_{0}^{2}\right) 
\label{eq:h:bro}
\end{align}
up to corrections in $\tau/T$, which are negligible if $T$ is large enough compared to the range of $\tau$ used. 
The entropy rate of the fBm is therefore linear in $\log(\tau)$, with a constant slope ${\cal H}$, independent of the temporal extension.  
We have computed the entropy rate $h_\tau^{(1)}$ of a synthesized fBM with ${\cal H}=1/3$ and containing the same number $N$ of points as our experimental data. 
Results are shown in Fig.\ref{fig:results:CMI}, the measured slope is $0.32\pm0.01$, in agreement with the theoretical value ${\cal H}=1/3$. We have also computed $h_\tau^{(m)}$ for $2\le m\le 4$ and observed a small deviation from the linear behavior for large $\tau$, while the slope slightly increases to reach $0.34\pm0.01$ for $m=4$. We attribute this to finite size effects, especially in the corrections to eq.(\ref{eq:h:bro}).

The fBm has no characteristic scales (no integral neither Kolmogorov scale), so the slope of its entropy rate is unperturbed by the integral and dissipative domains.
Therefore, analysing the behaviour of such a process with Hurst exponent $\mathcal{H}=1/3$ allowed us to explain the scaling behaviour of the entropy rate of our turbulence data in the inertial range.  

\section{Conditioned Entropy rate}

\begin{figure}
\centering
\includegraphics[width=1\linewidth]{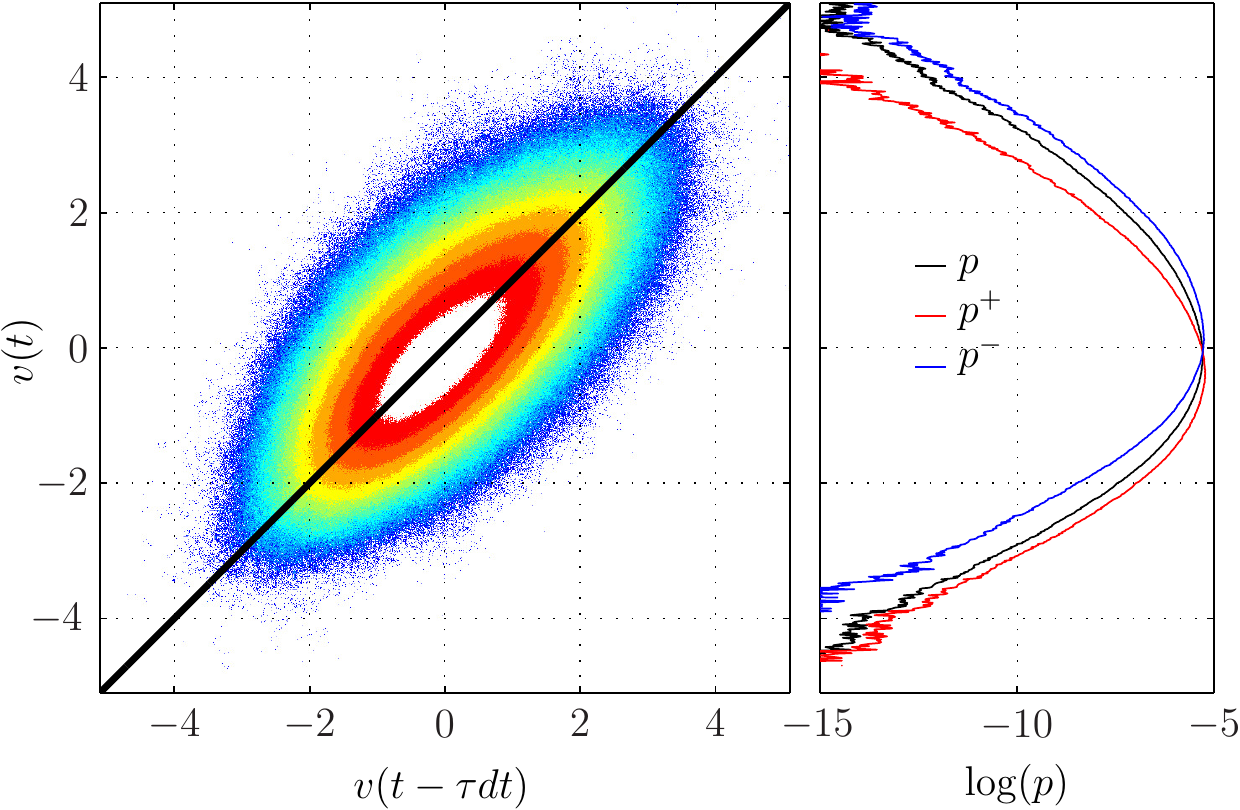}
\caption{Left: bi-variate PDF of  $(v(t),v(t-\tau dt))$, the velocity field.  
Right: $p(X)$ (in red), $p_+(X)$ (blue) and $p_-(X)$ (black).}
\label{fig:results:PDF}
\end{figure}

In the previous section we showed that the entropy rate of a Turbulence velocity signal is well described using only the autocorrelation function, because the signal statistics are very close to Gaussian. 
Nevertheless, the statistics of the increments of a Turbulent velocity signal $\delta_\tau(t) \equiv v(t)-v(t-\tau dt)$ can be far from Gaussian, especially for smaller scales $\tau$. In particular, the PDF of increments is skewed.
A non-vanishing skewness of the increments or a non-vanishing order-3 structure function, $S_3(r)\equiv\E \{(v(t)-v(t-\tau dt))^3\}$ results in an energy cascade~\cite{Kolmogorov1941a,Kolmogorov1941b,Kolmogorov1941c}.
The entropy rate, as defined in the former section, cannot probe the energy cascade because it is not sensitive enough to the asymmetry of the two-points PDF (represented in Fig.~\ref{fig:results:PDF}).

\subsection{Conditioned probabilities}
To probe more accurately the symmetry of the two points PDF of a generic signal $X$, we propose the following procedure. We define the signal $X_+$, resp. $X_-$, as the subset of points $x(t)$ from $X$ such that $\delta_\tau(t) >0$, resp. $\delta_\tau(t)<0$. We then define the conditioned PDF $p_+$, resp. $p_-$, of the signal $X_+$, resp. $X_-$;
Although signals $X_+$ and $X_-$ are one-dimensional and take their values in the same vector space $S$ as $X$ does, they contain some information provided by the increments, namely the sign of the local increment $\delta_\tau(t)$ associated with $x(t)$. 
It is important to note here that if the statistics of the increments $\delta_\tau(t)$ of the signal $X$ are skewed, then the joint PDF of $(x(t), x(t-\tau dt))$ is not symmetrical with respect to the origin. The reciprocal may not hold.
For the Turbulent velocity field, the conditioned PDFs are reported in Fig.~\ref{fig:results:PDF} for $\tau=400$.

\subsection{Conditioned entropy rate}
We then define the conditioned entropy rate $h_\tau^{+,(m)}$, resp. $h_\tau^{-,(m)}$, of the signal $X$ as the entropy rate $h_\tau^{(m)}$ of the conditioned signal $X_+$, resp. $X_-$. 
In practice, we compute the same quantity as before, defined by eq.(\ref{eq:def:h:MI}), but using only a subset of all data points; this subset is obtained by retaining points $x(t)$ with a given sign of $\delta_\tau(t)$.
The test is performed at the single date $t$, whatever the embedding dimension $m$ is, so not looking at the (sign of) increments $\delta_\tau(t-k\tau), 1\le k\le m-1$.

The entropy rate considers an embedded signal of dimension $m+1$, so even for the smallest $m=1$, the entropy rate probes 2-points correlations, between $x(t)$ and $x({t-\tau dt})$. In that case, the additional 
conditioning on the sign of the increment allows the conditioned entropy rate to probe the asymmetry of the joint PDF $p(x(t), x({t-\tau dt}))$, which is related to the skewness of the increments.

If the joint PDF is symmetrical with respect to the origin, it is easy to check that $p_+(x)=p_-(-x)$ and therefore
\begin{equation*}
h_\tau^{+,(m)}(X)=h_\tau^{-,(m)}(X).
\end{equation*}

On the contrary, if the statistics of the increments are skewed, then the joint pdf does not have the central symmetry and we may have $h_\tau^{+,(m)}(X)\neq h_\tau^{-,(m)}(X)$.

In the case of a fBm, the joint PDF $p(B_t, B_{t-\tau})$ has the central symmetry for all scales $\tau$. 
In the case of a Turbulence velocity signal, the joint PDF is not symmetrical for small scales $\tau$ because of the skewness of the increments.
\begin{figure}
\begin{center}
\includegraphics[width=1\linewidth]{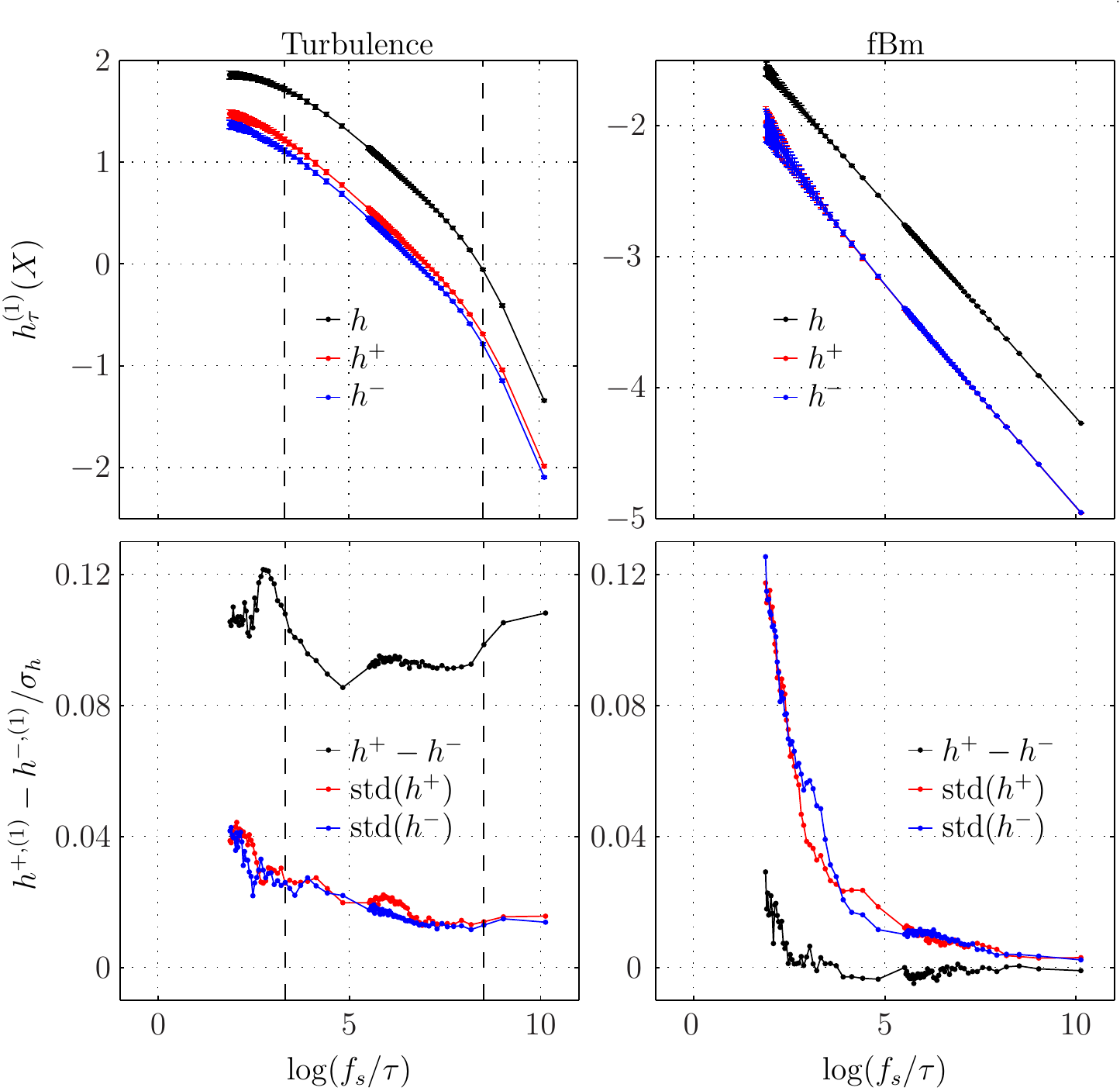}
\caption{Top: conditioned entropy rates $h_\tau^{+,(1)}$ (red) and $h_\tau^{-,(1)}$ (blue) as function of scale for Turbulence data (left) and fBm with $\mathcal{H}=1/3$ (right) for embedding $m=1$. The entropy rate $h^{(1)}_\tau$ (black) is reported for comparison.
Bottom: difference $(h_\tau^{+,(1)}-h_\tau^{+,(1)})$ (black) and standard deviation of $h_\tau^{+,(1)}$ (blue) and $h_\tau^{-,(1)}$ (red).} 
\label{fig:results:CMI} 
\end{center}
\end{figure}
Measurements of conditioned entropy rates are reported in Fig.\ref{fig:results:CMI} for $m=1$. 
For the fBm, the conditioned entropy rates are indistinguishable and follow the same linear behavior as the entropy rate $h_\tau^{(m=1)}$. 
For the turbulent velocity signal, the two conditioned entropy rate give significant different results, with an almost constant  difference around $0.1\pm0.01$ (compared to the standard deviation of $h_\tau^{(1)}(X^\pm)$
which is around $0.03\pm0.01$) . 

To check the robustness of our results, we perform two different tests using the Turbulence signal.
First, because of the skewness of the increments, especially for the smaller scales $\tau$, the fraction of points of $X$ in subsets $X^+$ and $X^-$ can be quite different (about 48\% and 52\% respectively). As this may cause a difference in conditioned entropy rates, we recomputed $h_\tau^{-,(1)}$ when imposing that $X^-$ has the same number of points as $X^+$. To do so, we simply discarded the extra points from $X^-$ (randomly chosen). This procedure does not change the correlations of $X^-$, nor the statistics and hence the pdf. Again, we find the same significant difference between $h_\tau^{+,(1)}$ and $h_\tau^{-,(1)}$.
Second, we replaced the conditioning on the sign of the increments by a random sub-sampling of $X$, in order to obtain a subsampled signal $X^{\text{rand}}$. In that case we obtain
$h_\tau^{(1)}(X^\text{rand})=h_\tau^{(1)}(X)$, as expected, although the number of points in $X$ and $X^{\text{rand}}$ differ by a factor 2.

\section{Discussion}

The entropy rate depends on the scale $\tau$ used in the embedding process. It measures the dynamics ---~understood here as the dependence on either temporal or spatial scales~--- of information in the signal at scale $\tau$. Using a particular value of $\tau$ for our estimates amounts to 
probe the signal at the scale $\tau$ or equivalently to use a signal down-sampled at the scale $\tau$. Because of the almost Gaussian nature of the statistics, we argued that the entropy rate is a function of the 
auto-correlation function ---  or equivalently of $S_2$~---- and it is therefore not surprising to observe three distinct regions in the "spectrum of information". 
Interestingly, the entropy rate discriminates the different regions ---~and hence the different
behaviors of the flow~---- differently than the Power Spectrum does. This is because the entropy rate is a non-linear function of the auto-correlation. 
The representation of the integral, inertial, and dissipative ranges given by the entropy rate for embedding $m=1$ contains the same physics as the one given by the second order structure function.
Both of them operate in direct space and so they avoid the Fourier transform that induces humps in the transition regions~\cite{Lohse1995}.

\medskip

We now comment on the values taken by the entropy rate. 
In the integral domain, $h^{(m)}_\tau$ is equal to the entropy $H(X)$ of the signal, which is independent of the scale $\tau$ and represents only the standard deviation $\sigma$ of the signal:
$$
\lim_{\tau \rightarrow \infty} h^{(m)}_\tau = H(X) \simeq \frac{1}{2} \log(2 \pi e \sigma^2)
$$
This is very generic and can also be understood as the absence of correlations in this domain, so that the MI term in eq.(\ref{eq:def:h:MI}) vanishes. Over the inertial range, the entropy rate decreases when 
decreasing the scale, indicating that the signal is more and more regular, or equivalently that a new point in the signal brings less and less information. The scaling of the inertial range (with a slope of $-1/3$) is exactly the one of a fBM or of 
the Batchelor model. In the dissipative range, the decrease is faster: with a slope of $-1$ for $m=1$, in accordance with Batchelor's model, and with a larger negative slope if $m>1$.

\medskip

Although the entropy rate depends {\em a priori} on the complete PDF of the signal, and therefore on higher order statistics, when $m=1$ we haven't been able to measure any significant deviation from Gaussianity: 
formula (\ref{eq:h:corr}) holds at all scales. Increasing the embedding dimension $m$ leads interestingly to a similar conclusion: whether one uses 2-point ($m=1$) or $n$-point statistics ($m=n-1$), the "Information 
Spectrum" is unchanged, in both the inertial and integral domains. Deviations are observed for the smallest scales only. There is no reason for this to hold for an arbitrary signal from a complex system, and we 
are currently investigating increments in more details, as well as other systems, to probe if this feature is characteristic of Turbulence.

\medskip

The conditioning procedure introduced to probe the skewness of increments ---~and hence the existence of an energy cascade~--- should be distinguished from the conventional "conditional entropy". 
Our conditioned entropy rate is not the conditional entropy rate, which is a Kullbach-Leiber divergence, but simply the entropy rate of a pre-conditioned signal. It allows us to probe very fine details of the 
statistics, and especially the asymmetry of the bivariate pdf which is related to the skewness of the increments.
We showed that the existence of an energy cascade can be proven using only Information Theory (and, of course, Kolmogorov 4/5 law).

The conditioning process may seem very crude, as we only test for the sign of the increments. We have tried to use the value of the increments as a condition, which amounts to compute the conditional entropy 
rate $H(X_{\tau}| X^{(m,\tau)},\delta_{\tau})$. Unfortunately this quantity is not sensitive to the skewness of the increments: all possible values of the increments are averaged over, which cancels the effect of the 
asymmetry of the bivariate PDF.

\section{Conclusion}

We measured the information content of a turbulent signal by computing the entropy rate as a function of the scale at which the signal is considered. We found that the distribution/scaling of information is reminiscent of the energy scaling, and we related it to the second order structure function. 
The entropy rate is able to separate properly the different domains as the second order structure function does.
We argued that the entropy rate is more sensitive to correlations than the Power Spectrum, in particular because it can take into account higher order correlations, especially for large embeddings.
For this reason, Information Theory perspective may give some new insight on Turbulence.

We reported that the entropy rate of the longitudinal velocity is unable to probe either the weak skewness of the Turbulence signal or the larger skewness of its increments. We then designed a conditioning 
procedure of the data, based on the sign of the increments. Applying the entropy rate to this conditioned data, we were able to illustrate an effect of the skewness of the velocity increments.

\medskip

The procedures described here are of general interest for the study of complex systems, especially those having multiscale dynamics, as can be found in, {\em e.g.}, Economy, Ecology, Neuroscience, and of course Fluid dynamics.
Given the plethora of laws governing such different systems, the model-free and nonlinear nature of Information Theory makes it a very interesting approach. In the particular case of Turbulence, we showed not only that 
the entropy rate allows one to measure in all generality the information distribution amongst scales, in perfect agreement with known models, but also that a well-chosen conditioning of the data allows one to prove the existence of the energy cascade.
Using Information Theory only, we recovered all classical characteristics of second order moment of Turbulence, as well as the existence of an energy cascade via the third order moment.

The authors wish to thank L. Chevillard for stimulating discussions.
This work was supported by the LABEX iMUST (ANR-10-LABX-0064) of Universit\'e de Lyon, within the program "Investissements d'Avenir" (ANR-11-IDEX-0007) operated by the French National Research Agency (ANR).

\bibliographystyle{unsrt}
\bibliography{biblio.bib}

\end{document}